\documentclass[aps,preprint,showpacs,superscriptaddress,groupedaddress]{revtex4}  % for double-spaced preprint
\usepackage{graphicx}  % needed for figures
\usepackage{dcolumn}   % needed for some tables
\usepackage{bm}        % for math
\usepackage{amssymb}   % for math
\usepackage{amsmath}
\usepackage[colorlinks=true, citecolor=blue, urlcolor=blue, linkcolor = blue]{hyperref}
\usepackage{bm}
\usepackage{color}
\usepackage{bookmark}
\usepackage{tabularx}
\usepackage{mathtools}
\usepackage{microtype}
\usepackage{cancel}
\usepackage{relsize}
\usepackage{float}
\usepackage[caption = false]{subfig}
\usepackage{physics}

\begin{document}

% The following information is for internal review, please remove them for submission
% \widetext
% \leftline{Version xx as of \today}
% \leftline{Primary authors: Joe E. Physics}
% \leftline{To be submitted to (PRL, PRD-RC, PRD, PLB; choose one.)}
% \leftline{Comment to {\tt d0-run2eb-nnn@fnal.gov} by xxx, yyy}
% \centerline{\em D\O\ INTERNAL DOCUMENT -- NOT FOR PUBLIC DISTRIBUTION}
% 
% % the following line is for submission, including submission to the arXiv!!
%\hspace{5.2in} \mbox{Fermilab-Pub-04/xxx-E}

\title{Correspondence between Dicke-model semiclasscial dynamics in the superradiant dipolar phase and the Euler heavy top}
%\input author_list.tex       % D0 authors (remove the first 3 lines
                             % of this file prior to submission, they
                             % contain a time stamp for the authorlist)
                             % (includes institutions and visitors)

%\affiliation{Instituut-Lorentz, Universiteit Leiden, P.O. Box 9506, 2300 RA Leiden, The Netherlands}

\author{S. I. Mukhin}
\affiliation{Theoretical Physics and Quantum Technologies Department, NUST ``MISIS'', Moscow, Russia}
\author{A. Mukherjee}
\affiliation{Theoretical Physics and Quantum Technologies Department, NUST ``MISIS'', Moscow, Russia}
\affiliation{Department of Physics, School of Basic Sciences, Manipal University Jaipur, Rajasthan, India}
\author{S.S. Seidov}
\affiliation{Theoretical Physics and Quantum Technologies Department, NUST ``MISIS'', Moscow, Russia}

\date{\today}

\begin{abstract}
 Analytic expression is found for the frequency dependence of transmission coefficient of a transmission line inductively coupled to the microwave cavity with superradiant condensate. Sharp transmission drops reflect condensate's frequencies spectrum. These results pave way to direct detection of emergence of the superradiant condensates in quantum metamaterials. Results are based on the analytic solutions of the nonlinear semiclassical dynamics of superradiant photonic condensate in the Dicke model of an ensemble of two-level atoms dipolar coupled  to the electromagnetic field in the microwave cavity. In adiabatic limit with respect to photon degree of freedom the system is approximately integrable, with evolution being expressed via Jacobi elliptic functions of real time. Depending on the coupling strength, the semiclassical coordinate of superradiant condensate in the ground state either oscillates in one of the two degenerate minima of condensate's potential energy or traverses between them over the saddle point. An experimental setup for measuring of the breakdown of the normal phase of the Dicke model via coupling to the transmission line is proposed. A one-to-one mapping of semiclassical motion of superradiant condensate on the nodding of unstable Lagrange "sleeping top" also turns Dicke model into analogue device for modelling dynamics of mechanical systems.  \end{abstract}

\maketitle

%\section{\label{sec:level1}First-level heading}
% sections are not used for PRL papers
\section{Introduction}
Prediction of  superradiant quantum phase transition \cite{Brandes, Brandes_entlg}, that breaks parity symmetry of the system consisting of  $N\gg1$ two-level (TL) atoms  coupled to a single bosonic mode in the resonant cavity, poses an interesting problem concerning observable fingerprints of  superradiant condensates emerging in the quantum metamaterials \cite{vonDelft, Mukhin, Fistul, Nakamura, Ciuti, Rabl}. This knowledge is also important for the quantum computation perspectives \cite{Wallraff,  Wallraff2, Raimond, DiCarlo}. Recently an approximate integrability of the Dicke model \cite{Dicke} was established \cite{EPL} in the adiabatic limit with respect to photon condensate degree of freedom. We show below, that in the vicinity of the quantum phase transition into superradiant state the condensate characteristic frequences obey the adiabaticity condition: $\Omega_n \ll \omega_0$, where $\hbar\omega_0$ is bare TL splitting. This allows analytic solution of the semiclassical dynamics equation for the superradiant condensate. This solution is based on the two integrals of motion possessed by the coupled photonic condensate and TL system in the adiabatic limit.  We found solutions that bear unexpected parallelism with the evolution of the polar angle made by the pivoted axis of Euler symmetric spinning top with direction of the external gravitational field, i.e. nodding of Lagrange 'sleeping top' \cite{Landau,Yagasaki, Murakami, Ashbaugh, Chao}. We demonstrate, that semiclassical dynamics of the superradiant condensate in the resonant cavity, and hence, of the Lagrange 'sleeping top' nodding, could be studied by measuring frequency dependence of the transmission coefficient of a transmission line inductively coupled to the cavity. This follows from the derived below analytic expression featuring sharp drops of transmission coefficient  at the characteristic frequencies of the superradiant condensate's spectrum. Our analytic solution indicates, that patterns of transmission coefficient drops along the frequencies axis depend on the coupling strength of TL system to the microwave cavity photons. 

\section{Dicke Hamiltonian in adiabatic approximation}
In this article we consider Dicke model Hamiltonian (DH) expressed in terms of  the operators of collective variables:
\begin{align}
\hat{H} =  \frac{\omega}{2}\left(\hat{p}^2+\hat{q}^2 \right) + \frac{2\gamma}{\sqrt{S}}\hat{q}\hat{S}^x  + \omega_0\, \hat{S}^z \label{UHU_CPB},
\end{align}
where $\hat{S}^\alpha=\sum_i \hat{s}_i^\alpha$ are Cartesian components of the total pseudo-spin  of  the TL system, and spin-$1/2$ Pauli operators $\hat{s}_i^\alpha$ characterise states of the $i$-th TL, the Planck constant $\hbar$ is taken for unity. We use here and below notations introduced in \cite{EPL}, where $\hbar=1$. The photon field second quantised operators are: 
\begin{align}
\hat{p}=\mathrm{i} \sqrt{\frac{1}{2}}\left(\hat{a}^\dag -\hat{a} \right)\;\;\; \text{and} \;\;\;  \hat{q}= \sqrt{\frac{1}{2}}\left(\hat{a}^\dag + \hat{a} \right)\, \label{pq},
\end{align}
where $\left[\hat{a},\hat{a}^\dag\right]=1$. The superradiant regime is achieved for $f^2 = \gamma^2/\gamma_c^2 > 1$, 
$\gamma_c=\sqrt{\omega\omega_0}/2$ \cite{Brandes}.
 Using then Born--Oppenheimer approximation for a slow semiclassical motion of the photonic condensate on the background of fast coherent TL systems  \cite{EPL} one may substitute operators $\hat{p}$ and $\hat{q}$ with c-numbers considered as fixed parameters with respect to the fast superspin degrees of freedom. Then the spin part in Eq. (\ref{UHU_CPB}) is diagonalized using rotation angle $\theta$ around $y$-axis (compare \cite{MukhinPRA, EPL}):
\begin{align}
&\hat{S}^z\cos{\theta}+\hat{S}^x\sin{\theta}=\hat{J}^z;\\
&\cos{\theta}=\frac{\omega_0}{\omega_P(q)};\,\sin{\theta}= \frac{2q\gamma}{\omega_P(q)\sqrt{S}} \label{theta};\\
&\omega_P(q)=\omega_0\sqrt{1+f^2\frac{q^2\omega}{S\omega_0}},
\label{diag}
\end{align} 
and consequently one obtains the Hamiltonian with adiabatic invariant $\hat{J}^z$:
\begin{align}
\hat{H}_a =  \frac{\omega}{2}\left({p}^2+{q}^2 \right) +\omega_P(q) \hat{J}^z  \label{H_diag}.
\end{align}
Since $\hat{H}_a$ commutes with $\hat{J}^z$, the lowest energy band is reached in the state $\ket{S,-S}$ with $\hat{J}^z\ket{S,-S}=-S\ket{S,-S}$.
Substituting this spin projection into Eq. (\ref{H_diag}) one finds effective Hamiltonian of the condensate:
\begin{align}
{H}_a(S,-S) =  \frac{\omega}{2}\left({p}^2+{q}^2 \right) -\omega_P(q)S  \label{const}.
\end{align}
Hence, Eq. (\ref{const}) describes a 'particle' moving in the potential:
\begin{equation}
U(q) = \frac{\omega}{2}q^2 - \omega_0 S \sqrt{1 + f^2\frac{\omega q^2}{\omega_0S}}.\label{Uq}
\end{equation}
Considering the first integral of motion of the Hamiltonian (\ref{const}) of a 'particle' with effective mass $1/\omega$ , one finds differential equation for the dynamic variable $q$:
\begin{align}
{H}_a(S,-S) =  \frac{\dot{q}^2}{2\omega}+U(q)=E  \label{diff}.
\end{align}
 Now we consider the case when the square root in Eq. (\ref{Uq}) can be expanded in powers of coordinate $q$, i.e. the following condition holds:
 \begin{equation}
 f^2\frac{\omega q^2}{\omega_0S}\ll 1.\label{small}
\end{equation}

Then, expanding the root in Eq. (\ref{Uq}) up to fourth order in $q$ and substituting it into Eq. (\ref{diff}), one obtains equation of motion of a 'particle'  in the double--well potential $U_\text{dw}$ (the constant term $-\omega_0 S$ is absorbed by constant $E$):
\begin{equation}
E = \frac{\dot q^2}{2 \omega} + \frac{\omega_0}{8S} \left(f^2\frac{\omega}{\omega_0}\right)^2 q^4 + \frac{\omega}{2}(1-f^2) q^2\equiv  \frac{\dot q^2}{2 \omega} +U_\text{dw}(q)
\label{eq_m}
\end{equation}
Condition of superradiance $f^2 > 1$ makes the last term negative, thus forming a double-well potential. 

\subsection{Applicability of the series expansion}
The minima $\pm q_\text{min}$ of $U_\text{dw}(q)$ are found readily from condition $\partial_q{U}_\text{dw}(q)|_{q_\text{min}}=0$. After substitution $q_\text{min}$ into applicability condition Eq. (\ref{small}) one obtains the following inequality condition:
\begin{equation}
 f^2\frac{\omega q_\text{min}^2}{\omega_0S}\equiv \frac{2(f^2-1)}{f^2}\ll 1.\label{mima}
\end{equation}
\noindent
Substituting $q_\text{min}$ into ${U}_\text{dw}(q)$ and using condition in Eq.(\ref{mima}) one finds:
\begin{equation}
\left|{U}_\text{dw}(q_\text{min})\right|\equiv \left|{{U}}_\text{dw}\right|_\text{min}= \dfrac{(f^2-1)^2\omega_0S}{2f^4}\ll \dfrac{\omega_0S}{8}.\label{mim}
\end{equation}
\noindent
Hence, approximate  polynomial expression in Eq. (\ref{eq_m}) for potential energy is safe to use in the close enough vicinity of the phase transition $f^2\rightarrow 1+0$. Therefore, allowing for the limitation Eq. (\ref{mima}), it seems at first glance, that the adiabaticity condition for the Hamiltonian  Eq. (\ref{H_diag}) would be $\omega\ll \omega_P$, which in our case of $f\approx 1$ could be achieved via inequality $\omega\ll \omega_0$, i.e. far from the resonance: $\omega=\omega_0$, compare \cite{EPL}. We shall see below, that this is not the case in the vicinity of the saddle-point energy $|E|\leq \left|{{U}}_{dw}\right|_{min}$, when adiabaticity is granted already by $f^2\rightarrow 1+0$ itself, even though the resonant condition $\omega_0 = \omega$ holds. 

Also one can calculate value of $\sin \theta$ in (\ref{theta}) at $q = q_\text{min}$, the result is
\begin{equation}
\sin\theta(q_\text{min}) = \sqrt{1 - \frac{\gamma_c^{4}}{\gamma^4}}\equiv\sqrt{1 - f^{-4}} . 
\end{equation}
Exactly this result was obtained for the superradiant phase in the Dicke model using the rotated Holstein--Primakoff transformation \cite{MukhinPRA}. Thus, $\theta$ is the "superradiant" angle, which describes (pseudo)spin rotation from $z$--axis to $x$--axis under the superradiant phase transition.

\subsection{Applicability of the adiabatic approximation}
As the system evolves and photonic coordinate $q(t)$ changes with time, there exists some probability of tunnelling to the upper energy band at $q=0$ where the gap between the $\ket{S, -S}$ and $\ket{S, -S + 1}$ bands is the smallest, due to Landau--Zener tunnelling. One can think about the problem in a following way: the spin subsystem is controlled by an external field $q(t)$ and its spectrum is changing in time as $q(t)$ changes in time. Original Landau--Zener problem considers the time--dependent Hamiltonian in form \cite{Zagoskin}
\begin{equation}\label{eq:H_LZ}
H_{LZ} = \begin{pmatrix}
\alpha t & \Delta\\
\Delta & -\alpha t
\end{pmatrix}.
\end{equation}
Then if, for example, the particle was initially in the ground state $\ket{g}$ with energy $-\sqrt{\alpha^2 t^2 + \Delta^2}$ at $t = -\infty$, the probability of it ending up in an exited state $\ket{e}$ with energy $\sqrt{\alpha^2 t^2 + \Delta^2}$ at $t = \infty$ (Landau--Zener tunnelling) is given by
\begin{equation}
P_{LZ} = e^{-\frac{\pi \Delta^2}{2\hbar |\alpha|}}.
\end{equation}  
From this expression follows, that the bigger is the gap $\Delta$ or smaller is the rate of change of the energy $\alpha$, the smaller is the tunnelling probability.

In our case, when considering tunnelling from spin state $\ket{J_z = -S}$ to state $\ket{J_z = - S + 1}$ due to change in time of $q(t)$, we restrict the whole phase space to only the considered subspace and the spin part of Hamiltonian (\ref{UHU_CPB}) is written as a two by two matrix
\begin{equation}
H_{S} = \begin{pmatrix}
\omega_0(-S + 1) & \sqrt{2} \gamma q(t) \\
\sqrt{2} \gamma q(t) & -\omega_0 S 
\end{pmatrix}.
\end{equation}
In order to bring it in form (\ref{eq:H_LZ}) we perform a unitary transformation 
\begin{equation}
\begin{aligned}
&H_{S} \rightarrow O^\dagger H_S O = 
\begin{pmatrix}
\sqrt{2} \gamma q(t) + \omega_0/2 & -\omega_0/2\\
-\omega_0/2& -\sqrt{2} \gamma q(t) + \omega_0/2
\end{pmatrix} - \omega_0 S\\
&O = \frac{1}{\sqrt{2}}
\begin{pmatrix}
1 & -1\\
1 & 1
\end{pmatrix}.
\end{aligned}
\end{equation}
The rate of change $\alpha$ is obtained by linearisation:
\begin{equation}
2 \gamma  q(t) \approx  2 \gamma \dot q(t_0) (t-t_0).
\end{equation}
Thus the rate $\alpha$ is $\alpha = 2 \gamma \dot q(t_0)$. Finally, the transition probability is 
\begin{equation}\label{eq:P_LZ}
P_{LZ} = \exp{-\frac{\pi \hbar^2 \omega_0^2}{|4 \hbar \gamma \dot q(t_0)|}}.
\end{equation}

The separation between non--adiabatic energy levels $\pm 2 \gamma q(t) + \omega_0/2$ is large far from the point $q=0$, thus the speed $\dot q(t)$ might not be small in this region. However, near $q=0$ the gap reduces and in order to suppress the tunnelling the rate  of change of $q(t)$ should be slow. The velocity $\dot q(t)$ at $q = 0$ can be easily found using energy conservation law (\ref{eq_m}):
\begin{equation}
\dot q \eval_{q = 0} = \sqrt{\frac{2\omega}{\hbar}[E - U_\text{dw}(0)]} = \sqrt{ \frac{2\omega}{\hbar} E}.
\end{equation}
The closer is the total energy to the maximum of the potential (\ref{eq_m}) at $E = 0$ from above, the less is the "particle's" speed when passing the maximum. And if the energy is below the maximum, i.e. $E < 0$, the "particle" does not even reach the point $q = 0$ at the maximum of the potential, so the tunnelling to the upper band is suppressed by large energy gap. Finally we express the Landau--Zener tunnelling probability via energy:
\begin{equation}\label{eq:P_LZ_E}
P_{LZ} \eval_{E > 0} = \exp{-\frac{\pi \hbar^2 \omega_0^2}{4 \gamma \sqrt{2\hbar \omega E}}}.
\end{equation} 
Later in Sec. \ref{sec:E_GS} we will calculate the Landau--Zener tunnelling probability for the energy $E$ equal to the ground state energy of the Dicke model.

\section{Solution in Jacobi functions}
Analytic solutions of equation of motion Eq. (\ref{eq_m}) in the quartic double well potential are well known \cite{wang} and found below in the form of the different Jacobi elliptic functions, depending on the value $E$ of the total energy of the system. Using conservation law (\ref{eq_m}) the motion of the photonic subsystem can be described by the differential equation
\begin{equation}
\frac{\dot q^2}{2 \omega} = E - \frac{\omega^2 f^4}{8 \omega_0 S}q^4 - \frac{\omega}{2}(1 - f^2) q^2.
\end{equation}

If the total energy of the system is positive: $E>0$, i.e. the 'particle' has enough momentum to surpass the potential barrier at $q=0$, the solution is:
\begin{equation}
\begin{aligned}
&q(t)_{E>0} = A \operatorname{cn}(\Omega t, k);\;A^2 = \frac{2S (f^2 - 1)}{f^4}\frac{\omega_0}{\omega}\left(1 + \sqrt{1 + \frac{2f^4 E}{S \omega_0 (1- f^2)^2}}\right)\\
&\Omega^2 = \omega^2(f^2 - 1)\sqrt{1 + \frac{2 f^4 E}{S \omega_0 (1-f^2)^2}};\;k^2 = \frac{1}{2}\left(1 + \frac{1}{\sqrt{1 + \frac{2 f^4 E}{S \omega_0 (1-f^2)^2}}}\right).\label{pos}
\end{aligned}
\end{equation} 
One should note, that $q(t)$ is a sign changing function, meaning that the 'particle' travels between the two wells $q=\pm q_0$.

For negative total energy, $E<0$, solution is:
\begin{equation}
\begin{aligned}
&q(t)_{E<0} = \pm A \operatorname{dn}(\Omega t, k);\;A^2 = \frac{\omega_0}{\omega}\frac{4S (1 - f^2) 2z}{f^4\left(1 - \sqrt{4z + 1}\right)}\\
&\Omega^2 = \frac{\omega^2(1-f^2)2z}{1 - \sqrt{4z + 1}};\;k^2 =\frac{\sqrt{4z + 1}}{2z}\left(\sqrt{4z + 1} - 1\right)\\
&z = \frac{E f^4}{2 \omega_0 S(1 - f^2)^2}.\label{neg}
\end{aligned}
\end{equation}
In this case $q(t)$ does not change sign, so that 'particle' remains in one of the two potential wells.

If the total energy is zero, $E=0$, the period of motion becomes infinite and solution is: 
\begin{equation}
q(t)_{E = 0} = \pm \frac{2\sqrt S}{f^2}\sqrt{\frac{\omega_0}{\omega}(f^2 - 1)}\sech(\omega t \sqrt{f^2 - 1}).\label{zer}
\end{equation}
This solution describes 'particle' moving infinitely long from the 'turning point' at $q=q(t=0)$ to the saddle point at $q(t=\infty)=0$.
In all the solutions above the inequality in Eq. (\ref{mima}) ensures adiabaticity condition $\Omega\ll \omega_0$ provided that $\omega\leq\omega_0$, and:
\begin{equation}
|E|\leq \left|{{U}}_{dw}\right|_{min}= \dfrac{(f^2-1)^2\omega_0S}{2f^4}.\label{adb}
\end{equation}
\noindent
Simultaneously, the semiclassical approximation is justified by the proportionality of the photonic condensate coordinate $q(t)\propto \sqrt{S}=\sqrt{N/2}$, where ${N}$ is the macroscopic number of the TL's.   

\section{Ground state energy of the Dicke model}\label{sec:E_GS}
\begin{figure}[h!!]
\subfloat[$f = 1.044$]{\includegraphics[width = 0.33\textwidth]{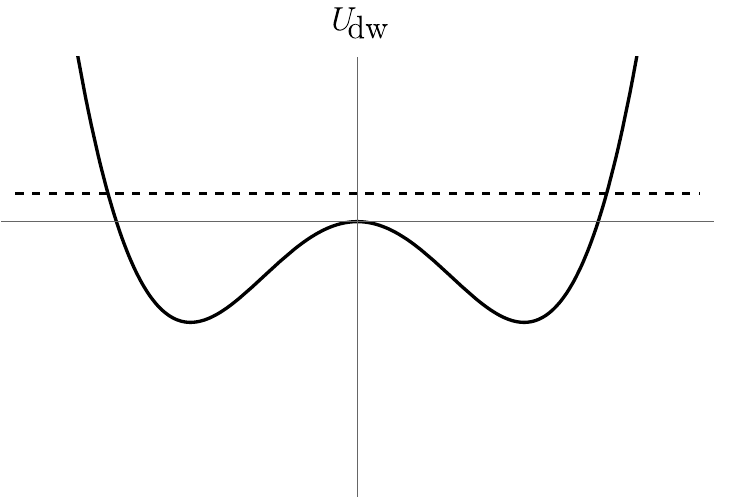}}
\subfloat[$f = 1.051$]{\includegraphics[width = 0.33\textwidth]{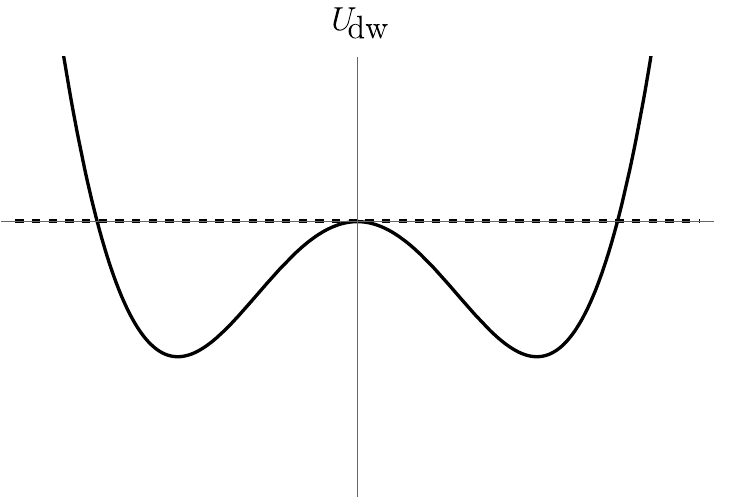}}
\subfloat[$f = 1.061$]{\includegraphics[width = 0.33\textwidth]{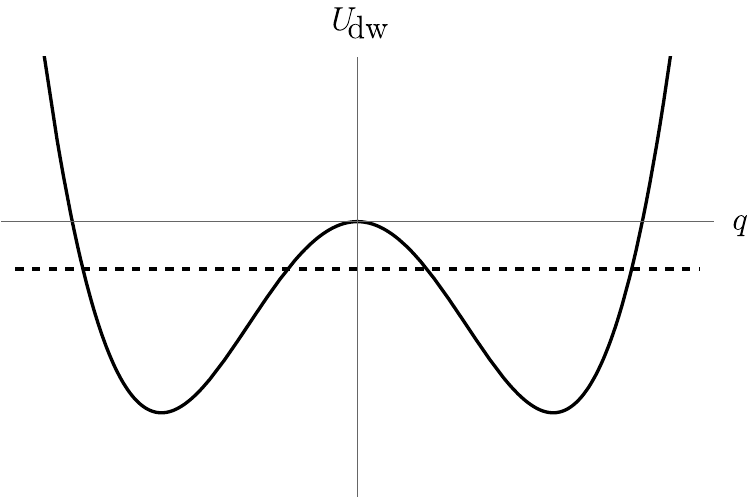}}
\caption{Illustration of the ground state energy level (dashed line) of the Dicke model with respect to the maximum of the effective potential in (\ref{eq_m}) (solid line). The parameters are chosen as  $\omega=1, \omega_0=5, S=10$. As parameter $f$ grows, the ground state energy level drops below  the potential energy maximum.}
\label{fig:E_GS}
\end{figure}

Our solutions above describe quasiclassical motion with arbitrary energy. However in quantum case the energy of course should be quantized and its arbitrary values are not allowed. In analogy with the Bohr model of electron in an atom, we consider discrete energy levels of the Dicke model as allowed values of energy of quasiclassical motion.  In particular, the analytical expression for the ground state energy $E_\text{GS}$ of the Dicke model in the superradiant phase can be obtained in the thermodynamic limit of large $S$ \cite{Brandes, MukhinPRA}, e.g. see equation (B15) in \cite{MukhinPRA}. In the notation, used in the current paper, the expression becomes 
\begin{equation}
\begin{aligned}
&E_\text{GS} = \omega_0 S - \frac{\omega_0 f^2}{2} (S + 1) - \frac{\omega_0}{2 f^2} S + \frac{1}{2}(\varepsilon_1 + \varepsilon_2)\\
&\varepsilon_{1,2}^2 = \frac{1}{2}\qty(\omega_0^2 f^4 + \omega^2 \pm \sqrt{\qty(\omega_0^2 f^4 - \omega^2)^2 +4 \omega^2 \omega_0^2}).
\end{aligned}
\end{equation}
This expression is valid for $f \geqslant 1$. Simultaneously, the value of the potential energy $U$ in (\ref{eq_m}) at its maximum equals zero. In particular, in the case of $f = 1$:
\begin{equation}\label{eq:E_GS_f1}
E_\text{GS} \eval_{f=1} = \frac{1}{2}\qty(\sqrt{\omega^2 + \omega_0^2} - \omega_0) > 0.
\end{equation}
This means, that at the superradiant quantum phase transition the ground state energy is above the potential barrier. Thus, the quasiclassical motion with total energy $E_\text{GS}$ in this regime is described by Jacobi $\operatorname{cn}$ function, see equation (\ref{pos}). As coupling constant grows further, at some value of $f>1$, the energy of the ground state drops below the top of potential barrier, i.e. $E_\text{GS} < 0$. This is illustrated in Fig. \ref{fig:E_GS}. By substituting (\ref{eq:E_GS_f1}) in (\ref{eq:P_LZ_E}) and in limit $\omega_0 \gg \omega$ we obtain the following expression for the probability of Landau--Zener tunneling for the quasiclassical dynamics with the energy, defined by the energy of the ground state of the quantum model at the point of the superradiant phase transition:
\begin{equation}
P = \exp \left\{ -\frac{\pi \omega_0^2}{\sqrt{2} \omega^2}\right\} .
\end{equation} 
This expression defines the upper bound for the probability of Landau--Zener tunnelling. The tunnelling is suppressed for $\omega_0 > \omega$, which is an agreement with the result for applicability of adiabatic approximation \cite{EPL} according to which the non--adiabatic coefficient
\begin{equation}\label{eq:C}
C = \langle S, -S; q| \hat p| S, -S + 1; q \rangle \bigg\rvert_{q = 0} = \frac{f}{2}\sqrt\frac{\omega}{\omega_0}
\end{equation}
should be small.

%%%%%%%%%%%%%%%%%
%%%%%%%%%%%%%%%%
\section{Measuring normal phase breakdown via transmission line coupled to photonic condensate}
\begin{figure}[h!!]
\centerline{\includegraphics[width = 0.72\linewidth]{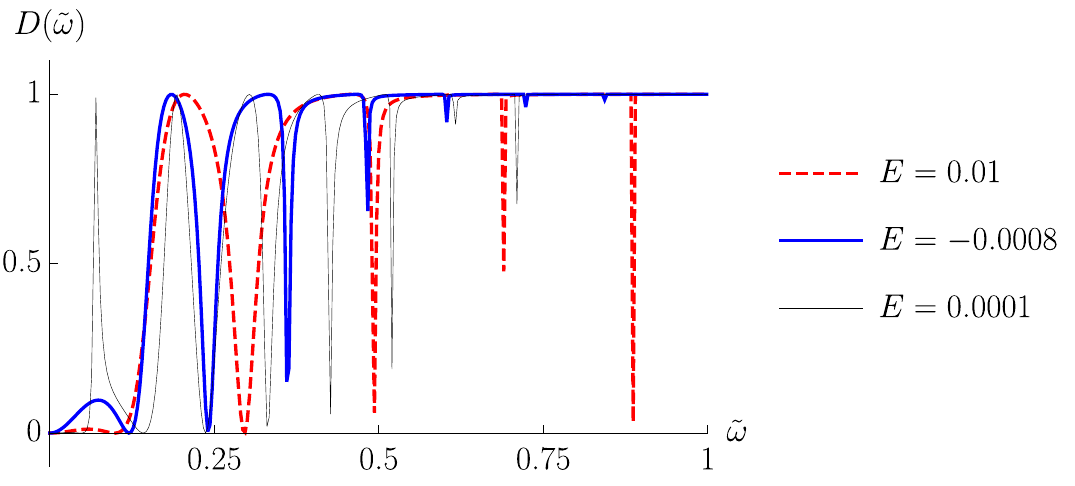}}
\caption{Frequency dependences of the transmission coefficient $D(\tilde\omega)$ Eq. (\ref{D}) for a transmission line inductively coupled to the microwave cavity with the superradiant condensate at different values of the coupling strength $f$, that enters  Eqs. (\ref{pos})-(\ref{neg}). Propagation velocity $c$ and coupling constant $\kappa$ in  Eq. (\ref{semi}) obey relation: $c \kappa^2 = 0.0225$. Other parameters are the same as in Fig. \ref{fig:E_p}.}
\label{fig:Domega}
\end{figure}

The normal phase of the Dicke model would correspond to the case of a single-well potential in Eq. (\ref{eq_m}), when $f^2 < 1$ and $q=0$ at the minimum of the effective potential energy. The emergence of the superradiant condensate at $f^2 > 1$ causes transition to a bifurcating equilibria, i.e. to the double-well potentials in Eq. (\ref{eq_m}), thus, making the normal phase unstable. Simultaneously, the semiclassical motions of the photonic condensate Eq. (\ref{pos}) and Eq. (\ref{neg}) could be measured e.g. using transmission line setup \cite{Fistul}. In this setup an electromagnetic wave $Q(y,t)$ propagating in the transmission line (along axis $y$) is described by the Maxwell propagation equation of an electromagnetic wave coupled linearly to the superradiant condensate via  semiclassical coordinate $q(t)$: 

\begin{eqnarray}
&\dfrac{\partial^2Q(y,t)}{c^2\partial t^2}- \dfrac{\partial^2Q(y,t)}{\partial y^2}=\kappa\delta(y-y_0)q(t) \label{line}\\
&H={H}_a(S,-S)-\kappa q(t)Q(y_0,t),
\label{semi}
\end{eqnarray}
\noindent
where $c$ is the electromagnetic wave propagation velocity and $\kappa$ is the strength of inductive coupling  between the transmission line and the microwave cavity, that contains condensate described by the unperturbed Hamiltonian Eq. (\ref{diff}). Considering now the last term in Eq. (\ref{semi}) as a perturbation, one finds a response $q_1(t)$ of the condensate, linear in $\kappa Q(y_0,t)$, that follows from Eq. (\ref{eq_m}):

\begin{equation}
 \frac{\ddot q_1}{\omega} +\dfrac{\partial^2 U_{dw}(q_0)}{\partial q^2}q_1\equiv \hat{L}q_1=\kappa Q(y_0,t),
\label{resp}
\end{equation}
\noindent where unperturbed  condensate motions $q_0(t)$, expressed in Eqs. (\ref{pos}) - (\ref{zer}), obey Eq. (\ref{eq_m}). Allowing for the latter,  one finds that Eq. (\ref{resp}) is of Lam\'e type, with 'external force' $\kappa Q(y_0,t)$ and, hence, its solution looks like:

\begin{eqnarray}
 &q_1(t)=\int_{0}^{t}G^R(t-t')\kappa Q(y_0,t')dt' \label{q1}\\
 &\hat{L}G^R(t-t')=\delta(t-t');\;\;G^R(t-t'<0)\equiv  0, \label{Gdef}
\end{eqnarray}
\noindent where retarded Green's function $G^R$ of the Lam\'e differential operator $\hat{L}$ is introduced. Substituting Eq. (\ref{q1}) into right-hand side of the Maxwell's equation Eq. (\ref{line}) and making its Fourier transform with respect to time variable $t$, one finds a Schr\"odinger like equation for a scattering from the Dirac delta function potential barrier, which results in a transmission coefficient $D(\tilde\omega)$ of the transmission line for a wave travelling with the frequency $\tilde\omega$  \cite{Flugge}:

\begin{eqnarray}
&-\partial^2_yQ_{\tilde\omega}-\kappa^2G^R(\tilde\omega)\delta(y-y_0)Q_{\tilde\omega}=\dfrac{\tilde\omega^2}{c^2}Q_{\tilde\omega};\label{schrod}\\
&D(\tilde\omega)=\left |\dfrac{2\tilde\omega}{2\tilde\omega-ic\kappa^2G^R(\tilde\omega)}\right |^2 \label{D}.
\end{eqnarray}
\noindent The Green's function $G^R(t)$ is easily constructed allowing for the fact, that zero modes of Lam\'e  differential operator $\hat{L}\partial_t{q}_0=0$ are just the first time-derivatives of the corresponding unperturbed solutions, ${q}_0(t)$, of the Hamiltonian dynamics Eq. (\ref{eq_m}) expressed via  the Jacobi elliptic functions in Eqs. (\ref{pos})-(\ref{neg}):

\begin{eqnarray}
G^R(t\geq 0)=\begin{cases}-\dfrac{\omega}{\Omega^2}\partial_t{\operatorname{cn}}(\Omega t, k);\; & E>0\\
-\dfrac{\omega}{\Omega^2}\partial_t{\operatorname{dn}}(\Omega t, k);\; & E<0;
 \end{cases} \label{GQ};\;G^R(t<0)\equiv 0.
\end{eqnarray}
\noindent Using Fourier series expansions of  the Jacobi elliptic functions \cite{wang} ${\operatorname{cn}}(\Omega t, k), {\operatorname{dn}}(\Omega t, k) $ one finds $G^R(\tilde\omega)$ as the sum of the Green's functions $G_{\Omega_n}^{0R}(\tilde\omega)$ of harmonic oscillators with frequencies forming (half)integer multiples of photonic condensate frequency $\pi\Omega/{K}$:

\begin{eqnarray}
&G_{E>0}^R(\tilde\omega)=\dfrac{\pi^2 \omega}{2\Omega kK^2}\displaystyle\sum_{n=0}^{\infty}\dfrac{(2n+1)G_{\Omega_n}^{0R}(\tilde\omega)}{\operatorname{ch}[{(2n+1)\pi K'}/{2K}]},\;\Omega_n=\dfrac{(2n+1)\pi\Omega}{2K}\,; \label{cn}\\
&G_{E<0}^R(\tilde\omega)=\dfrac{\pi^2 \omega}{\Omega K^2}\displaystyle\sum_{n=1}^{\infty}\dfrac{nG_{\tilde{\Omega}_n}^{0R}(\tilde\omega)}{\operatorname{ch}[{n\pi K'}/{K}]},\; \tilde{\Omega}_n=\dfrac{n\pi\Omega}{K}\;,\label{dn}\\
&G_{\Omega}^{0R}(\tilde\omega)=\dfrac{1}{2}\left[\dfrac{1}{\tilde\omega+\Omega+i\delta}-\dfrac{1}{\tilde\omega-\Omega+i\delta}\right]\;,\; K'=K(\sqrt{1-k^2}).\label{G0}
\end{eqnarray}
\noindent Corresponding frequency dependences of the transmission coefficient $D(\tilde\omega)$ for the transmission line are presented in Fig. \ref{fig:Domega}. The presence of the superradiant condensate is reflected by the sharp transmission drops at the frequencies belonging to the  Fourier spectrum of the semiclassical motion of the condensate, the latter being marked by the poles of the Green's function in Eqs. (\ref{cn}), (\ref{dn}).

\noindent Relative narrowing of the intervals between transmission coefficient sharp drops in Fig. \ref{fig:Domega} when one moves from $E=0.01$ to  $E=0.0001$, i.e. more close to the saddle-point energy of photonic condensate $E=0$, is remarkable. It is related with the fact that solutions for $k$-values in Eq. (\ref{pos}) and Eq. (\ref{neg}) provide the limit $k\rightarrow 1$ when $E\rightarrow 0$. Since the complete elliptic integral of the first kind $K(k\rightarrow 1)\rightarrow\infty$ \cite{wang} the frequencies spectrum $\Omega_n$ of the superradiant condensate given in Eqs. (\ref{cn}), (\ref{dn}), condenses in the direction of  zero frequency. 

\section{Mapping on dynamics of spinning Lagrange top}
\begin{figure}[h!!]
\centerline{\includegraphics[width = 0.35\linewidth]{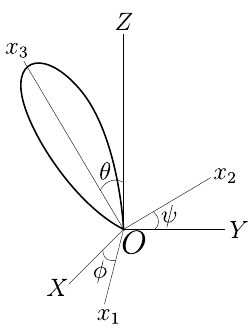}}
\caption{Scheme of the Euler top with pivoted point of the $x_3$-axis and the definition of the Euler angles.}
\label{Top}
\end{figure}

Using definition of $\theta$-angle in Eq. (\ref{theta}) in combination with equation of motion of the photon field coordinate $q$ in Eq. (\ref{eq_m}) in the limit Eq. (\ref{small}), it is straightforward to find a striking coincidence of the photonic coordinate dynamics with the dynamics of the Euler $\theta$-angle of a symmetric top in gravitational field \cite{Landau}, the Lagrange top. For this purpose we remind this fundamental problem in classical mechanics. It is well known that the Euler-Poisson equations of motion of a rigid symmetric top, that moves about a fixed point under the action of a gravitational force (the Lagrange case), are integrable \cite{Landau}. The energy conservation law for this case is written as: 
\begin{equation}
\tilde{E}=\frac{\tilde{I}_1}{2}(\dot{\theta}^2+\dot{\phi}^2\sin^2\theta)+\frac{{I}_3}{2}(\dot{\psi}+\dot{\phi}\cos\theta)^2+\mu gl\cos\theta.
\label{eudau}
\end{equation} 
\noindent Here the Euler angles $\theta, \phi, \psi $ are introduced as usual \cite{Landau}, relative to the laboratory $x,y,z$ coordinate system, and gravity acts in the negative direction of the polar $z$-axis direction, see Fig. \ref{Top}. 
\noindent
Then, angle $\theta$ is formed by the $z$-axis and the axis $x_3$ of the rotating top, i.e. one of its three principal axes $x_{1,2,3}$ , with the corresponding moments of inertia $I_1=I_2, I_3$ .  The pivoting point and the centre of mass both lie on the $x_3$-axis, separated by the distance $l$. 
\begin{figure}[h!!]
\subfloat[a]{\includegraphics[width=0.45\linewidth]{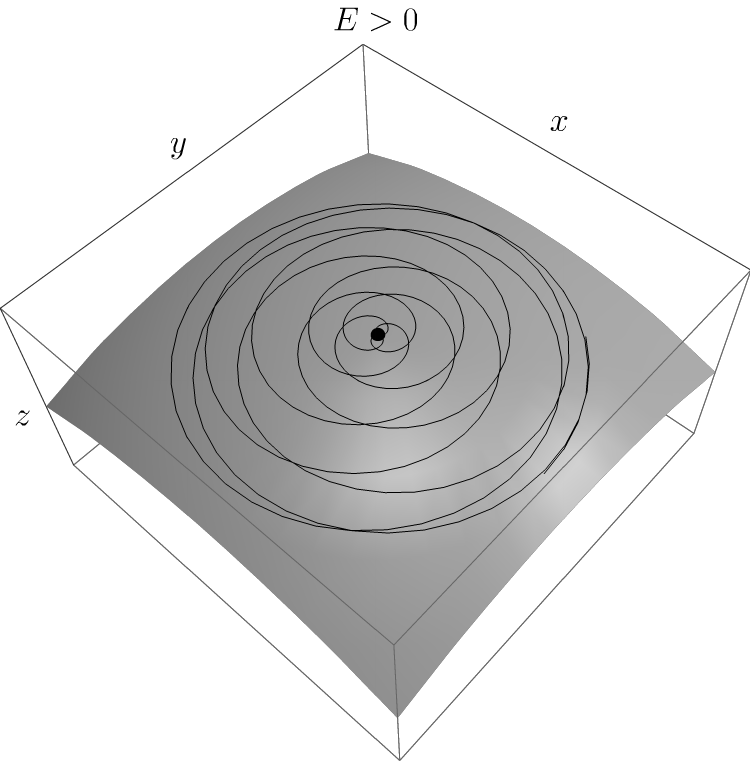}}\hspace{0.2cm}
\subfloat[b]{\includegraphics[width=0.45\linewidth]{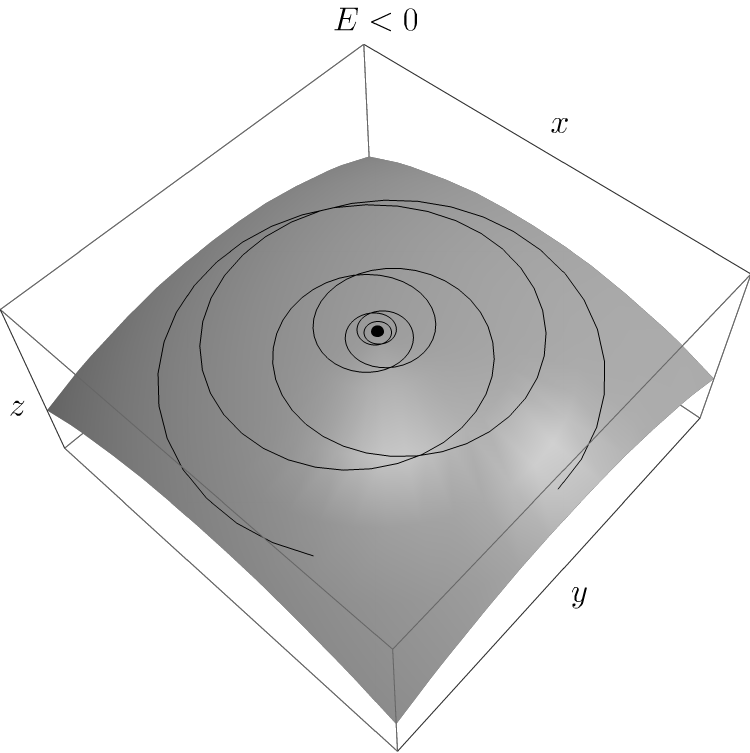}}\\
\subfloat[c]{\includegraphics[width=0.45\linewidth]{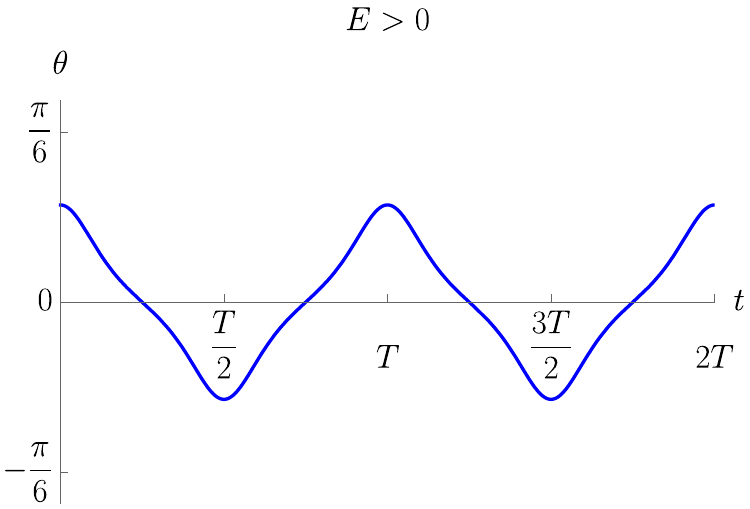}}\hspace{0.2cm}
\subfloat[d]{\includegraphics[width=0.45\linewidth]{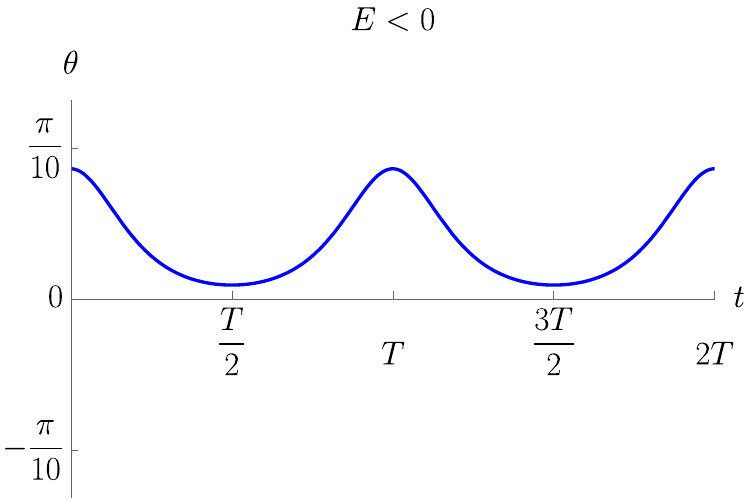}}
\caption{Trajectories on the sphere of the free end of the $x_3$-axis of  the spinning Lagrange top with: a) positive energy $E=0.01$ and $f = 1.21$, b) negative energy $E = -0.03$ and $f = 1.22$. Other parameters of the Dicke model Hamiltonian Eq. (\ref{UHU_CPB}): $\omega=1, \omega_0=5, S=10$. Corresponding time-evolution of the angle $\theta(t)$, c) and d), is taken from Dicke model solutions Eq. (\ref{pos}), Eq. (\ref{neg}) and definition Eq. (\ref{theta}) of the angle $\theta$. $T$ is the oscillation time period of the corresponding Jacobi function.}
\label{fig:E_p}
\end{figure}

\noindent
Constants $\mu$ and $g$ are the mass  and the acceleration of the top in the gravitational field, and $\tilde{I}_1\equiv I_1+\mu l^2$. Since the conjugate momenta corresponding to the cyclic angles $\phi, \psi $ are conserved, one obtains the following extra integrals of motion:
\begin{eqnarray}
&p_{\psi}=I_3(\dot{\psi}+\dot{\phi}\cos\theta)=M_3 \label{eudau21}\\
&p_{\phi}=(\tilde{I}_1\sin^2\theta+I_3\cos^2\theta)\dot{\phi}+I_3\dot{\psi}\cos\theta=M_z,
\label{eudau2}
\end{eqnarray} 
\noindent where integrals of motion $M_3, M_z$ are the angular moments of the top along the axes  $x_3$ and $z$ respectively, counted with respect to the fixed point $O$ of the top. Now, excluding
variables   $\dot \phi, \dot \psi $ from expression in Eq. (\ref{eudau}) using Eqs. (\ref{eudau21}), (\ref{eudau2}), and considering a particular case $M_3= M_z\equiv M$, one finds instead of Eq. (\ref{eudau}):
\begin{equation}
\begin{aligned}
&\tilde{E}-\dfrac{M^2}{2I_3}-\mu gl=\dfrac{\tilde{I}_1}{2}\dot{\theta}^2+\dfrac{M^2}{2\tilde{I}_1}\tan^2\dfrac{\theta}{2}-2\mu gl\sin^2\dfrac{\theta}{2}\approx\\
&\approx 2\tilde{I}_1\dot{Q}^2+\dfrac{M^2}{2\tilde{I}_1}\left[Q^4+Q^2\left(1-\dfrac{4\mu gl\tilde{I}_1}{M^2}\right)\right],
\end{aligned}\label{eudau3}
\end{equation}
\begin{equation}
\dot{\phi}=\dfrac{M}{2\tilde{I}_1(1-Q^2)},\; \dot{\psi}=\dfrac{M(I_3+2(\tilde{I}_1-I_3)(1-Q^2)))}{2\tilde{I}_1I_3(1-Q^2)},
\label{eudau4}
\end{equation}
\noindent where smallness of the Euler angle $\theta$ is assumed: $\sin\theta/2\equiv Q \ll 1$. Now, a direct comparison of Eq. (\ref{eudau3}) with Eq. (\ref{eq_m}), allowing for the inequality Eq. (\ref{small}), leads to the conclusion that the pseudospin rotation angle $\theta$ defined in Eq. (\ref{theta}) possesses the same dynamics as the Euler angle of the Lagrange top $\theta$, that enters dynamics equation Eq. (\ref{eudau3}). Namely, the one-to-one correspondence between equations Eq. (\ref{eq_m}) and Eq. (\ref{eudau3}) is achieved under the following conditions:
\begin{eqnarray}
{\mu gl}/{\tilde{I}_1}=\omega^2(2f^2-1),\;\;{M}/{2{\tilde{I}_1}}={\omega}{f}^{-1},\;\;\dfrac{E}{2\omega_0S}=\left[\tilde{E}-\dfrac{M^2}{2I_3}-\mu gl\right]\cdot({M^2}/2\tilde{I}_1)^{-1}.
\label{eudick}
\end{eqnarray}
\noindent
The analytical solutions Eq. (\ref{pos}) and Eq. (\ref{neg}) expressed via the angle $\theta$, as defined in Eq. (\ref{theta}), are plotted in the form of  trajectories of the spinning Lagrange top ( 'sleeping top' noddings) on the sphere, see Fig. \ref{fig:E_p}. The Euler angle $\phi$ depends linearly on time $t$ according to Eqs. ({\ref{eudau4}}), (\ref{eudick}):  
$\phi\approx\omega f^{-1} t$. This time dependence of the rotation angle $\phi$ around $Z$-axis was considered as fast one with respect to $\theta(t)$ and averaging over $\phi(t)$ was made in the equations of motion in \cite{Bound_luminosity}. 

\section{Conclusions}
In conclusion, we have considered the Dicke model in a superradiant state and found analytic solutions of  the semiclassical dynamics of the photonic condensate in the adiabatic limit with respect to the bare cavity mode coupled to the two-level atomic array. We have also discovered one-to-one correspondence between photonic condensate dynamics and nodding motion of  the Lagrange 'sleeping top'. Finally, we found analytic expression for the frequency dependence of the transmission coefficient of a transmission line inductively coupled to the microwave cavity with the superradiant photonic condensate. Predicted sharp transmission drops reflect Fourier spectrum of the semiclassical motion of photonic condensate and of a nodding 'sleeping top'. This opens way to observe directly the fingerprints of  photonic condensates emerging in the quantum metamaterials, as well as to use QED circuits for a simulation of dynamics of the classical mechanical systems. 
 
\section{ACKNOWLEDGMENTS}
The authors acknowledge illuminating discussion with Prof. Alexandre Zagoskin in the final stage of this work. The study was supported by the Federal academic leadership program Priority 2030 (NUST MISIS grant No. K2-2022-025).


\begin{thebibliography}{99}


\bibitem{Brandes} Clive Emary and Tobias Brandes, {Chaos and the quantum phase transition in the Dicke model}, \href{https://doi.org/10.1103/PhysRevE.67.066203}{Phys. Rev. E \textbf{67}, 066203 (2003)}.

\bibitem{Brandes_entlg} K. Hepp and E.H. Lieb, {On the Superradiant Phase Transition for Molecules in a Quantized Radiation Field:
the Dicke Maser Model}, \href{https://doi.org/10.1016/0003-4916(73)90039-0} {Ann. Phys. \textbf{76}, 360 (1973)}.

 
\bibitem{vonDelft} Oliver Viehmann, Jan von Delft, and Florian Marquardt, {\it Superradiant Phase Transitions and the Standard Description of Circuit QED}, \href{https://doi.org/10.1103/PhysRevLett.107.113602}{Phys. Rev. Lett. \textbf{107}, 113602 (2011)}.

\bibitem{Mukhin} S. I. Mukhin and M. V. Fistul, {Generation of non-classical photon states in superconducting quantum metamaterials}, \href{https://iopscience.iop.org/article/10.1088/0953-2048/26/8/084003/meta}{Supercond. Sci. Technol. \textbf{26}, 084003 (2013)}.

\bibitem{Fistul} M. A. Iontsev, S. I. Mukhin, and M. V. Fistul, {Double-resonance response of a superconducting quantum metamaterial:
Manifestation of nonclassical states of photons}, \href{https://doi.org/10.1103/PhysRevB.94.174510}{Phys. Rev. B \textbf{94}, 174510 (2016)}.

\bibitem{Nakamura} Motoaki Bamba, Kunihiro Inomata, and Yasunobu Nakamura, {Superradiant Phase Transition in a Superconducting Circuit in Thermal Equilibrium}, \href{https://doi.org/10.1103/PhysRevLett.117.173601}{Phys. Rev. Lett. \textbf{117}, 173601 (2016)}.


\bibitem{Ciuti} Pierre Nataf and Cristiano Ciuti,
{\it No-go theorem for superradiant quantum phase transitions in cavity QED and counter-example in circuit QED}, \href{https://doi.org/10.1038/ncomms1069}{Nature Communications \textbf{1}, 72 (2010)}.

\bibitem{Rabl} Tuomas Jaako, Ze-Liang Xiang, Juan José Garcia-Ripoll, and Peter Rabl, {Ultrastrong-coupling phenomena beyond the Dicke model}, \href{https://doi.org/10.1103/PhysRevA.94.033850}{Phys. Rev. A \textbf{94}, 033850 (2016)}.

\bibitem{Wallraff} Alexandre Blais, Ren-Shou Huang, Andreas Wallraff, S. M. Girvin, and R. J. Schoelkopf, {Cavity quantum electrodynamics for superconducting electrical circuits: An architecture for quantum computation}, \href{https://doi.org/10.1103/PhysRevA.69.062320}{Phys. Rev. A \textbf{69}, 062320 (2004)}.

\bibitem{Wallraff2} A. Wallraff, D. I. Schuster, A. Blais, L. Frunzio, R.-S. Huang, J. Majer, S. Kumar, S. M. Girvin, and R. J. Schoelkopf, {\it Strong coupling of a single photon to a superconducting qubit using circuit quantum electrodynamics}, \href{https://doi.org/10.1038/nature02851}{Nature \textbf{431}, 162-167 (2004)}.

\bibitem{Raimond} J. M. Raimond, M. Brune, and S. Haroche, {Manipulating quantum entanglement with atoms and photons in a cavity}, \href{https://doi.org/10.1103/RevModPhys.73.565}{Rev. Mod. Phys. \textbf{73}, 565 (2001)}.

\bibitem{DiCarlo} L. DiCarlo, M. D. Reed, L. Sun, B. R. Johnson, J. M. Chow, J. M. Gambetta, L. Frunzio, S. M. Girvin, M. H. Devoret, and R. J. Schoelkopf, {Preparation and measurement of three-qubit entanglement in a superconducting circuit}, \href{https://doi.org/10.1038/nature09416}{Nature \textbf{467}, 574-578 (2010).} 

\bibitem{Dicke} R. H. Dicke, {Coherence in Spontaneous Radiation Processes}, \href{https://doi.org/10.1103/PhysRev.93.99}{Phys. Rev. \textbf{93}, 99 (1954)}.

\bibitem{EPL}A. Relano, M. A. Bastarrachea-Magnani, and S. Lerma-Hernandez, { Approximated integrability of the Dicke model},\href{https://iopscience.iop.org/article/10.1209/0295-5075/116/50005}{EPL,  \textbf{116}, 50005 (2016)}. 

\bibitem{Landau} L. D Landau and E.M. Lifshitz, Mechanics: Volume 1 (Course of Theoretical Physics) 3rd Edition, Butterworth-Heinemann, 1976 , 170p.

\bibitem{Yagasaki} G. H. M. van der Heijden and Kazuyuki Yagasaki, {Horseshoes for the nearly symmetric heavy top}, Z. Angew. Math. Phys. \textbf{65}, 221-240 (2014).

\bibitem{Murakami} H. Murakami, O. Rios, T. J. Impelluso,  {A Theoretical and Numerical
Study of the Dzhanibekov and Tennis Racket Phenomena}, Journal of Applied Mechanics, \textbf{83}, 111006 ( 2016)

\bibitem{Ashbaugh} Ashbaugh, M., Chicone, C. C., and Cushman, A. H., The Twisting
Tennis Racket, J. Dyn. Differ. Equations, \textbf{3}(1), pp. 67-85 (1991).

\bibitem{Chao} Chao B., Feynman, Dining Hall Dynamics, Physics Today, \textbf{42}(2), 15 (1989).

\bibitem{MukhinPRA} S. I. Mukhin and N. V. Gnezdilov, First-order dipolar phase transition in the Dicke model with infinitely coordinated frustrating interaction,  {Phys.Rev.A, \textbf{97}, 053809 (2018)}.

\bibitem{Zagoskin} A. M. Zagoskin, Quantum Engineering: Theory and Design of Quantum Coherent Structures, 1st ed, Cambridge University Press (2011).

\bibitem{wang} Wang Z.X., Guo D.R., Special Functions, World Scientific Publishing Co Pte Ltd., 1989, 695p.  

\bibitem{Flugge}S. Fl\"ugge, Practical Quantum Mechanics, vol. 1, Springer, 1994, 331p.

\bibitem{Bound_luminosity} Mukhin S. I., A. Mukherjee, and S. S. Seidov, Dicke Model Semiclassical Dynamics in Superradiant Dipolar Phase in the “Bound Luminosity” State, {Journal of Experimental and Theoretical Physics 132, \textbf{4} (apr 2021)}

\end{thebibliography}
\end{document}